\documentclass[12pt,preprint]{aastex}

\newcommand{\beq}{\begin{equation}}
\newcommand{\eeq}{\end{equation}}

\newcommand{\kms}{km s$^{-1}$~}

\slugcomment{to appear in Ap. J. Letters}

\shortauthors{Benedict et al.}
\shorttitle{Mass of Gl 876b}

\received{2002 September 24}
\begin{document}

\title{A Mass for the Extrasolar Planet Gl 876b Determined from {\it Hubble Space Telescope} Fine Guidance Sensor 3 Astrometry and High-Precision Radial Velocities\footnote{Based on 
observations made with
the NASA/ESA Hubble Space Telescope, obtained at the Space Telescope
Science Institute, which is operated by the
Association of Universities for Research in Astronomy, Inc., under NASA
contract NAS5-26555}} 

\author{ G. F. Benedict\altaffilmark{1}, B. E. McArthur\altaffilmark{1}, T. Forveille {\altaffilmark{2}$^{\ ,}$\altaffilmark{3}}, X. Delfosse\altaffilmark{3}, E. Nelan \altaffilmark{4}, R. P. Butler\altaffilmark{5}, W. Spiesman\altaffilmark{1}, G. Marcy\altaffilmark{6}, B. Goldman\altaffilmark{8}, C. Perrier\altaffilmark{3}, W. H. Jefferys\altaffilmark{7}, and M. Mayor\altaffilmark{9} }
\altaffiltext{1}{McDonald Observatory, University of Texas, 1 University Station C1402, Austin, TX 78712, fritz @astro.as.utexas.edu, mca@astro.as.utexas.edu}
\altaffiltext{2}{ Canada-France-Hawaii Telescope, P.O. Box 1597, Kamuela, HI 96743, forveill@cfht.hawaii.edu }
\altaffiltext{3}{Observatoire de Grenoble, B.P. 53X, 38041 Grenoble Cedex, France, delfosse@obs.ujf-grenoble.fr,perrier@obs.ujf-grenoble.fr }
\altaffiltext{4}{Space Telescope Science Institute, 3700 San Martin Drive, Baltimore, MD 21218, nelan@stsci.edu}
\altaffiltext{5}{Department of Terrestrial Magnetism, Carnegie Institution of Washington, 5241 Broad Branch Road NW, Washington, DC 20015, paul@dtm.ciw.edu }
\altaffiltext{6}{Department of Astronomy, University of California, Berkeley, 601 Campbell Hall, Berkeley, CA 94720, gmarcy@etoile.berkeley.edu }
\altaffiltext{7}{ Astronomy Dept., University of Texas, 1 University Station C1400, Austin, TX 78712, bill@astro.as.utexas.edu }
\altaffiltext{8}{Department of Astronomy, New Mexico State University, Department 4500, P.O. Box 30001, Las Cruces, NM 88011, bgoldman@nmsu.edu }
\altaffiltext{9}{Observatoire de Genve, 51 Ch. des Maillettes, 1290 Sauverny, Switzerland, Michel.Mayor@obs.unige.ch }



\begin{abstract}
We report the first astrometrically determined mass of an extrasolar planet, a companion previously detected by Doppler spectroscopy. Radial velocities first provided an ephemeris with which to schedule a significant fraction of the {\it HST} observations near companion peri- and apastron. The astrometry residuals at these orbital phases exhibit a systematic deviation consistent with a perturbation due to a planetary mass companion. Combining {\it HST} astrometry with radial velocities, we solve for the proper motion, parallax, perturbation size, inclination, and position angle of the line of nodes, while constraining period, velocity amplitude, longitude of periastron, and eccentricity to values determined from radial velocities. We find a perturbation  semi-major axis and inclination, $\alpha$ = 0.25 $\pm$ 0.06 mas, $i$ = 84\arcdeg $\pm$6\arcdeg, and Gl 876 absolute parallax, $\pi_{abs}= 214.6 \pm$ 0.2 mas.  Assuming that the mass of the primary star is $M_* = 0.32M_{\sun}$, we find the mass of the planet, Gl 876b, $M_b = 1.89\pm0.34M_{Jup}$. 

\end{abstract}




%

\section{Introduction}
In 1998 two groups searching for unseen companions to stars using the technique of Doppler spectroscopy announced the discovery of a companion to the M4 dwarf star, Ross 780 = Gl 876 (Delfosse et al. 1998, Marcy et al. 1998).  This companion was characterized by $Msini \sim 2 M_{Jup}$ and P$\sim$60$^d$. Later, a second companion was detected, Gl 876c (Marcy et al. 2001), with $Msini=0.56 M_{Jup}$ and P$\sim$30$^d$. We were granted {\it HST} time to characterize the perturbation due to the longer period companion, GL 876b. The low mass of the primary ($M_* = 0.32M_{\sun}$), the period of Gl 876b, and the system proximity (4.7pc) suggested that even an edge-on orientation ($M = Msini$) would produce a detectable perturbation, thus a companion mass. A transiting companion of HD 209458 allowed  Henry et al. (2000) to establish the first mass of an extrasolar planet. We report the first {\em astrometrically} determined mass of an extrasolar planet.

\section{Observations and Reductions} \label{OR}

Observations
 were  obtained with FGS 3 in fringe tracking mode. The FGS instrument is described in Nelan \& Makidon (2002)\nocite{Nel02}.
Each of our 27 data sets required $\sim40$ minutes of spacecraft time. Data were reduced and calibrated as detailed in Benedict et al. (2002b)\nocite{Ben02b}, McArthur et al. (2001)\nocite{McA01}, and Benedict et al. (1999)\nocite{Ben99}. While acquiring each set, we measured reference stars and the Gl 876 multiple times to correct for intra-orbit drift (e.g. Benedict et al. 2002a\nocite{Ben02a}, Fig. 1). Three prime epochs (1, 3, and 4, boldface in Table~\ref{tbl-OBS}) contain 30 or 45 observations of Gl 876 and were obtained near companion peri- and apastron, as predicted from the radial velocity (RV) data. For a planetary mass companion of unknown inclination observations with HST/FGS precision at any other orbital phase are less likely to yield a detection. To minimize the contribution to our error budget from the FGS Optical Field Angle Distortion calibration (Benedict et al. 1999)\nocite{Ben99}, we held the spacecraft orientation fixed for the three prime epochs as shown in Table~\ref{tbl-OBS}. The last three data sets (epochs $7-9$ at a different spacecraft roll) were obtained significantly after the bulk of the observations. They maximize the signatures of stellar parallax and proper motion of Gl 876, so that these systematic motions could be precisely modeled. Table~\ref{tbl-OBS} contains mean epochs for the nine distinct observation campaigns. 

We obtained classification spectra (27 October 2000) of our astrometric reference stars with the European Southern Observatory Faint Object Spectrograph and Camera (EFOSC) on the 3.6m telescope. This effort yields a more accurate absolute parallax for Gl 876, reducing the contribution of distance uncertainty to the companion mass determination. Reference stars were classified by a combination of template matching and line ratios. The half-fringe produced by the FGS walk down to fine-lock (Nelan \& Makidon 2002\nocite{Nel02}) indicated that ref-4 is a binary star with $\Delta V\sim1$. With a system brightness, V = 11.02, the primary has an intrinsic V = 11.5. Absolute magnitudes for these spectral types from Cox (2000), V magnitudes from FGS photometry, and an adopted upper limit A$_V$ = 0.1 from Schlegel et al.~(1998)\nocite{Sch98} at the galactic latitude of Gl 876, produced the absolute parallaxes listed in Table~\ref{tbl-SPP}. These enter our modeling as observations with error.

With the positions measured by FGS 3 we determine the scale, rotation, and offset plate
constants for
each observation set relative to an arbitrarily adopted constraint epoch. The Gl 876 reference frame contains 5 stars. We employ a six parameter model and neglect corrections for lateral color discussed in Benedict et al. (1999), because the orientation was held constant for the epochs that were expected to yield the largest perturbation signals. 

As in all our previous astrometric analyses, we employ GaussFit (Jefferys, Fitzpatrick, \& McArthur 1987\nocite{Jef87}) to minimize $\chi^2$. The solved equations
of condition for the Gl 876 field are:

\beq
\xi  =  Ax + By + C + R_x (x^2 + y^2) - \mu_x \Delta t  - P_\alpha\pi_x
\eeq
\beq
\eta  =  -Bx + Ay + F + R_y(x^2 + y^2) - \mu_y \Delta t  - P_\delta\pi_y
\eeq

A, and B
are scale and rotation plate constants, C and F are
offsets; $R_x$ and $R_y$ are radial terms;
$\mu_x$ and $\mu_y$ are proper motions; $\Delta$t is the epoch difference from the mean epoch;
$P_\alpha$ and $P_\delta$ are parallax factors;  and $\it \pi_x$ and $\it \pi_y$
 are  the parallaxes in RA and Dec. We obtain the parallax factors from a JPL Earth orbit predictor (Standish 1990\nocite{Sta90}), upgraded to version DE405. Orientation to the sky is obtained from ground-based astrometry 
(USNO-A2.0 catalog, Monet 1998\nocite{Mon98}) with uncertainties of $ \pm  0\fdg05$.

We treat reference star proper motions (from the UCAC catalog, Zacharias et al. 2000\nocite{Zac00}) as observations with error. Results are shown in Table~\ref{tbl-POS}.
The standard errors resulting from the solutions for relative position, parallax,
and the final adjusted proper motions of Gl 876 and the reference stars
are $\sim 1$ milliarcsecond (mas). The resulting reference frame `catalog' in $\xi$ and $\eta$ standard coordinates transformed to RA and DEC relative to reference star ref-4  was determined
with average uncertainties $<\sigma_\xi> = 0.7$	 and	$<\sigma_\eta> = 0.5$ mas. Figure~\ref{fig-HISTO}
shows histograms of target and reference star residuals
obtained from our astrometric modeling, prior to including the perturbation in the model. 

\section{The Astrometric Detection of the Gl 876 Perturbation} \label{AST-only}

Once proper motion and parallax have been determined, we search for any remaining systematic patterns in the Gl 876 residuals. After transforming all residuals to an RA-Dec orientation, we plot them (Figure~\ref{fig-RAW}) phased to a period of Gl 876b as determined by RV measurements \cite{Mar01}, P = 61.02 days. Having phased the bulk of the astrometric observations to occur near peri- and apastron, the perturbation appears as a position difference between these two orbital phases in RA and Dec (Figure~\ref{fig-RAW}). The straight lines are least-squares fits to all the residuals in both RA and Dec, where the slopes in RA and Dec are 5$\sigma$ and 1$\sigma$ results, respectively. These slopes are determined almost entirely by the clumps of residuals at $\phi$ = 0.26 (periastron) and $\phi$ = 0.72. By definition, the total vector separation between the residual clumps at peri- and apastron is twice the perturbation semi-major axis, $2\alpha$. Hence, from astrometry alone we obtain $\alpha = 0.3 \pm 0.1$ mas. As in any traditional binary star measurement the photocentric orbit as measured by FGS 3 can be influenced by the brightness of the secondary (Benedict et al. 2001\nocite{Ben01}). Our simulations and past measurements of actual binary stars indicate that $\Delta m = 5$ is the largest $\Delta m$ that might shift the primary photocenter by an amount equal to the astrometric precision of our final semi-major axis result discussed in Section~\ref{RVAST}.  Assuming this $\Delta m$ and any of the lower main sequence mass-luminosity relations referenced below, the system distance, period, and primary mass would predict a 54 mas perturbation. As evidenced by the small perturbation, the companion must have $\Delta m >> 5$. The barycentric and photocentric position of Gl 876 are identical within our observational precision. 

\section{The Gl 876b Perturbation Size, Inclination, and Mass from Combined Astrometry and Radial Velocities} \label{RVAST}
Considering only astrometry at peri- and apastron (the two densest clumps seen in Figure~\ref{fig-RAW}), we could not distinguish between the inclination and orientation of the perturbation. Properly, our modeling considers the astrometry at all orbital phases. Our analysis techniques for defining orbits by combining {\it HST} astrometry and RV 
have been described  by Benedict et. al. (2001). There we determined all reference and target star parameters (including the traditional orbital elements) in a simultaneous solution incorporating both astrometry and RV. For Gl 876 we also employ techniques to combine RV from different sources (Hatzes et al. 2000) and, additionally, apply several constraints. First, we
constrain all plate constants to those determined from our astrometry-only solution (equations 1 and 2, Section~\ref{OR}). Second, 
we constrain the orbital elements K, e, P, and $\omega$ to those determined by us (simultaneously, for both the inner and outer companion) from all available RV data. The final constraint is the same one applied in Benedict et al. (2001)\nocite{Ben01}, the relationship between astrometric and RV determined parameters discussed in Pourbaix \& Jorissen (2000).
\beq
\displaystyle{{\alpha~sin~i \over \pi_{abs}} = {P K_1 sqrt(1 -e^2)\over(2\pi)\times4.7405}} 
\eeq
\noindent where quantities derived only from our astrometry (parallax, $\pi_{abs}$, primary perturbation size, $\alpha$, and inclination, $i$) are on the left, and quantities derivable for the outermost companion from RV (the period, $P$, eccentricity, $e$, and the RV amplitude for the primary, $K_1$), are on the right.

With RV measurements complete through 2001 (Keck, Lick, CORALIE, ELODIE), all astrometric measurements, and the constraints as described, we solve for parallax, proper motion, and the semi-major axis, orbit orientation, and orbit inclination for the perturbation caused by the outer companion. The model is aware of the existence of the inner companion, and the RV are treated as a sum of the velocities due to both Gl 876b and Gl 876c. But, for now the astrometric signature is treated as a perturbation due only to Gl 876b. For the parameters critical in determining the mass of Gl 876b we find a parallax, $\pi_{abs} = 214.6 \pm 0.2$ mas, a perturbation size, $\alpha = 0.25 \pm 0.06$ mas, and an inclination, $i = 84\arcdeg \pm 6$\arcdeg. These, proper motion, and the other constrained orbital elements are listed in Tables~\ref{tbl-AST} and ~\ref{tbl-ORB}, along with our mass for Gl 876b. The planetary mass critically depends on the mass of the primary star, for which we have adopted $M_* = 0.32\pm0.05M_{\sun}$ (Henry \& McCarthy 1993\nocite{Hen93}, Delfosse et al. 2000\nocite{Del00}, Henry \& Torres 2002\nocite{Hen02}). The first value in Table~\ref{tbl-ORB}, $M_b = 1.89\pm0.34M_{Jup}$, assumes no uncertainty in the mass of the primary star. The second value, $M_b = 1.9\pm0.5M_{Jup}$, incorporates the present uncertainty in $M_*$. Table~~\ref{tbl-AST} also shows satisfactory agreement for the parallax and proper motion of Gl 876 from {\it HST}, {\it HIPPARCOS} (HIP 113020), and the Yale Parallax Catalog (YPC95). 
If co-planar with Gl 876b, Gl 876c will produce a perturbation of order 0.05 mas.

\begin{figure}
\epsscale{0.75}
\plotone{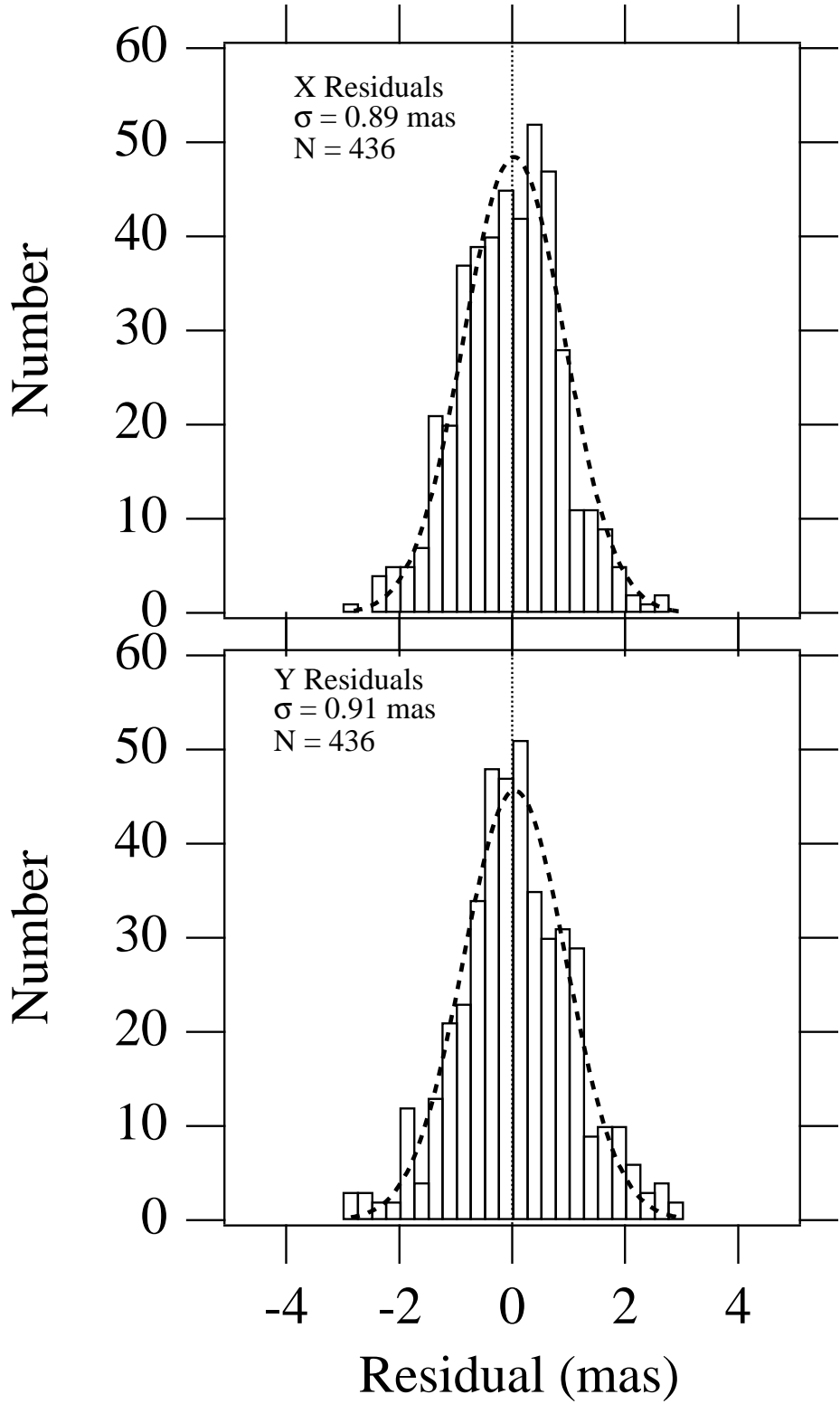}
\caption{Histograms of x and y astrometric residuals obtained from modeling  Gl 876 and reference stars with equations 1 and 2. Fits are 
Gaussians with indicated standard deviations.} \label{fig-HISTO}
\end{figure}

\begin{figure}
\epsscale{0.75}
\plotone{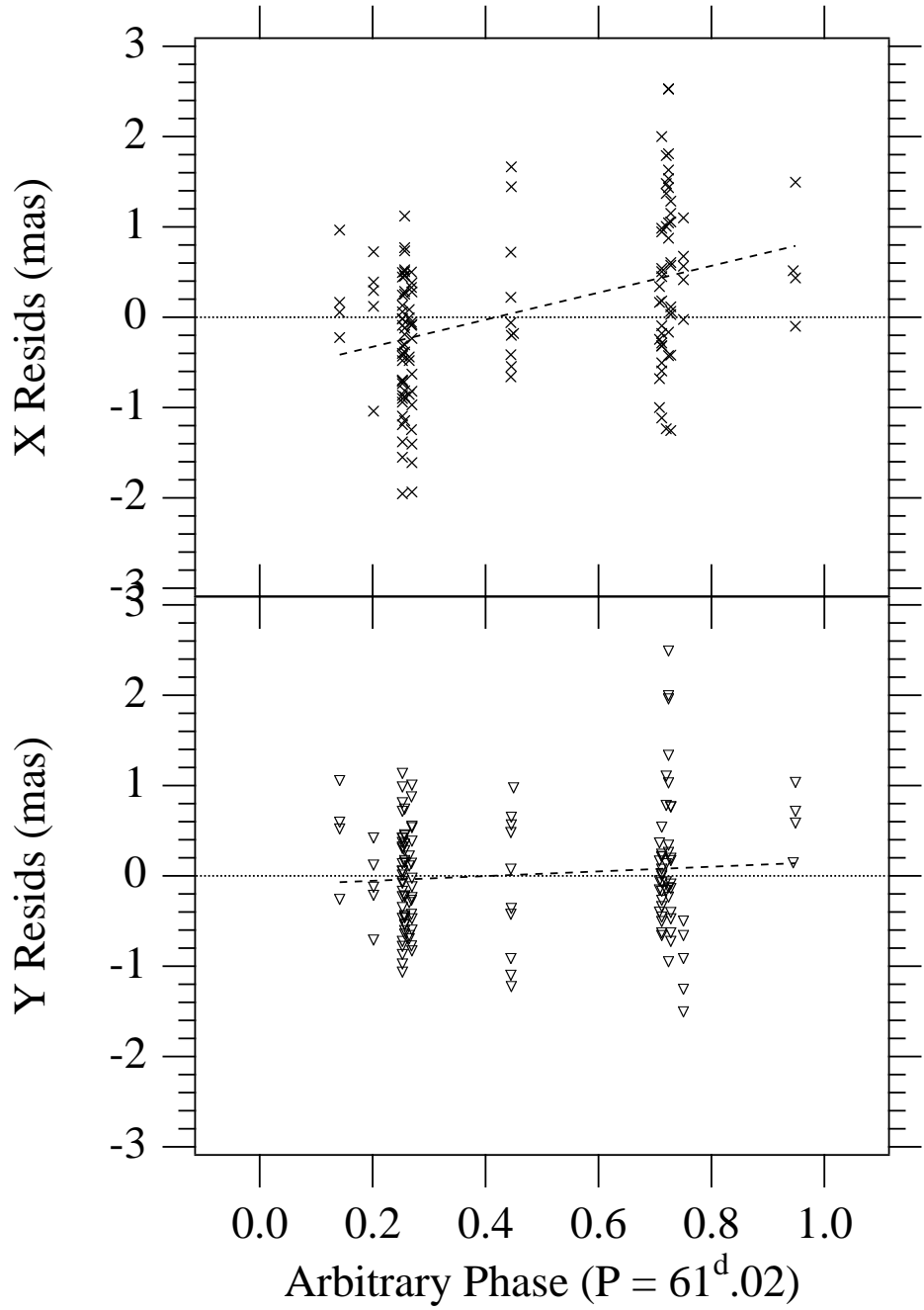}
\caption{Raw residuals for Gl 876 phased to the Gl 876b companion period, P = 61.02 days, determined from radial velocities. The offset between the two primary phases indicates the detection of a perturbation.} \label{fig-RAW}
\end{figure}

\begin{figure}
\epsscale{1.0}
\plotone{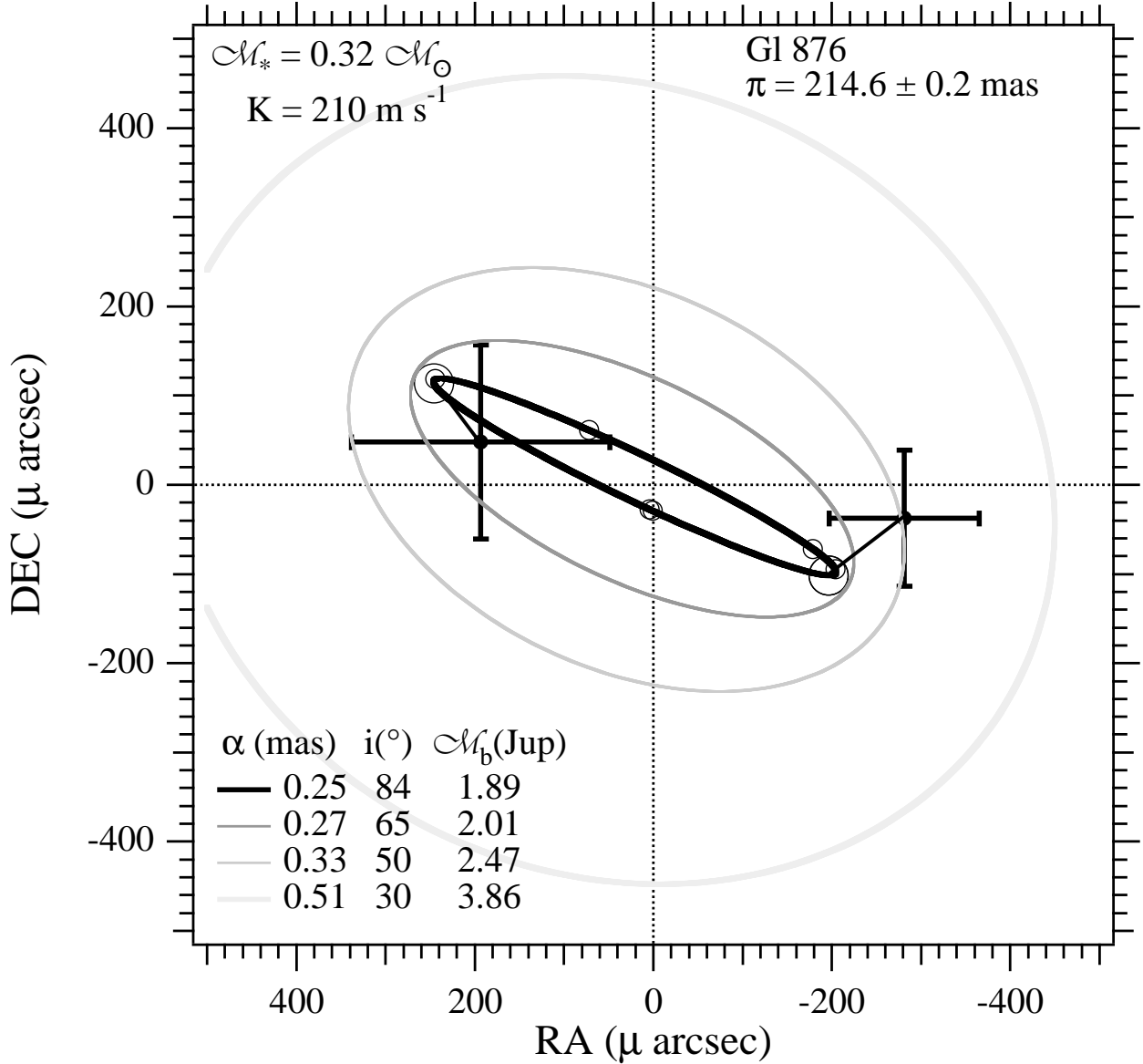}
\caption{Four orbits permitted by the Pourbaix \& Jorissen constraint (equation 3) for the perturbation of due to Gl 876b. The innermost orbit is determined from the simultaneous astrometry-RV solution discussed in Section 4, and is described by the orbital parameters in Table~\ref{tbl-ORB}. Plotted on this orbit are the phases of the astrometric observations, with the two primary phases indicated by large circles. Also plotted ($+$) are the astrometry-only residual normal points at $\phi$ = 0.26 (periastron, lower right) and $\phi$ = 0.72. These normal points are connected to the derived orbit by residual vectors. The insert table shows the variation of inclination as a function of perturbation size as mandated by equation 3. Changing inclination changes companion mass.} \label{fig-ORBS}
\end{figure}

To confirm the validity of our combined RV and astrometric modeling, we form two normal points with the residuals from the data-rich epochs 1, 3, and 4 (Table~\ref{tbl-OBS}).  We can collapse the clouds of residuals seen in Figure~\ref{fig-RAW} to normal points, because the astrometric residuals are distributed normally (Figure~\ref{fig-HISTO}). These two best-determined residuals at peri- and apastron are plotted in Figure~\ref{fig-ORBS}. The errors are standard deviations of the means on each axis. Comparing with four of the many possible orbits permitted by the equation 3 constraint, we find satisfactory consistency with the innermost orbit produced by our derived parameters (Table~\ref{tbl-ORB}). 

We have yet to modify our model to treat the mutual interactions of the two components, which are expected to result in time-varying orbital elements (Laughlin \& Chambers 2001\nocite{Lau01}, Rivera \& Lissauer 2001,\nocite{Riv01} Ji, Li, \& Liu 2002\nocite{Ji02}, Nauenberg 2002\nocite{Nau02}). We expect the effect of these changes to be small, because the perturbation magnitude is established using observations spanning only a single orbital period. As a check we input to our model the dynamically determined RV values of Laughlin \& Chambers (2001) and obtained a companion mass equal to that found by our dual Keplerian approach. We will eventually include evolution of orbital elements, which may further reduce the mass uncertainty for Gl 876b.

\acknowledgments

GFB, BEM, and WHJ received support through HST NASA grants
G0-08102 and GO-09233 (T. Forveille, P.I.).
We thank Otto Franz, Tom Harrison, and Denise Taylor for useful discussions, and an anonymous referee for clarifying suggestions.

\clearpage
\begin{center}
\begin{deluxetable}{lllll}
\tablewidth{0pt}
\tablecaption{{Gl 876 Log of Observations}
\label{tbl-OBS}}
\tablehead{
\colhead{Epoch} 
& \colhead{JD-2400000} 
& \colhead{$\phi$\tablenotemark{a}} 
&\colhead{N$_{obs}$\tablenotemark{b}} 
& \colhead{{\it HST} Roll\tablenotemark{c}}  
}
\startdata
{\bf 1}&{\bf 51333.6447}&{\bf0.259}&{\bf 30}&{\bf295.87}\\
2&51345.1173&0.447&5&295.87\\
{\bf 3}&{\bf 51361.6886}&{\bf0.719}&{\bf45}&{\bf295.87}\\
{\bf 4}&{\bf51394.6221}&{\bf0.259}&{\bf 30}&{\bf295.87}\\
5&51405.9592&0.444&5&295.87\\
6&51696.292&0.202&5&295.87\\
7&51851.7516&0.75&5&98.005\\
8&52229.8838&0.947&4&114.281\\
9&52241.8424&0.143&4&112.886\\
\enddata
\tablenotetext{a}{Arbitrary Gl 876b orbital phase, assuming P = 61.02 
days.}
\tablenotetext{b}{Number of Gl 876 observations at this epoch}
\tablenotetext{c}{FGS 3 orientation }
\end{deluxetable}
\end{center}

\begin{center}
\begin{deluxetable}{llllll}
\tablewidth{0pt}
\tablecaption{Astrometric Reference Stars: Spectral Classifications and
Spectrophotometric Parallaxes\label{tbl-SPP}}
\tablehead{
\colhead{ID}
& \colhead{SpT}
&\colhead{V\tablenotemark{a}} 
& \colhead{M$_V$\tablenotemark{b}} 
& \colhead{A$_V$\tablenotemark{c}}  
&\colhead{$\pi_{abs}$ (mas)}
}
\startdata
ref-2&G5 V&14.47&5.1$\pm$0.4&0.1&1.4$\pm$0.3\\
ref-3&G8 V&13.90&5.5$\pm$0.4&0.1&2.2$\pm$0.4\\
ref-4&G0 V&11.50\tablenotemark{d}&4.4$\pm$0.4&0.1&3.8$\pm$0.7\\
ref-5&K2 V&13.41&6.4$\pm$0.4&0.1&4.2$\pm$0.8\\
ref-6&K1 V&15.12&6.2$\pm$0.4&0.1&1.8$\pm$0.3\\
\enddata
\tablenotetext{a}{From FGS photometry reduced and calibrated as per Benedict et 
al. (1998)}
\tablenotetext{b}{From Cox (2000)}
\tablenotetext{c}{From Schlegel et al.~(1998)}
\tablenotetext{d}{After removing contribution from a stellar companion with $\Delta V = +1.0$}

\end{deluxetable}
\end{center}

\begin{deluxetable}{crrrrr}
\tablecaption{Gl 876 and Reference Star Astrometry \label{tbl-POS}
}
\tablewidth{0pt}
\tablehead{
\colhead{Star} & 
\colhead{$RA$ (arcsec)\tablenotemark{a}$^{\ ,}$\tablenotemark{b}}   &
 \colhead{$DEC$ (arcsec)\tablenotemark{a}$^{\ ,}$\tablenotemark{b}} &
\colhead{$\pi_{abs}$ (mas)\tablenotemark{b}}&\colhead{$\mu_{RA}$ 
(mas)\tablenotemark{b}}   &
 \colhead{$\mu_{DEC}$ (mas)\tablenotemark{b} }
}
\startdata
Gl 876&262.9840$\pm$0.0005&95.3963$\pm$0.0003&214.6$\pm$0.6& 954.1$\pm$0.8&-
674.4$\pm$0.6\\
ref-2&174.9630$\pm$0.0005&93.9411$\pm$0.0005&1.4$\pm$0.1 &5.2$\pm$0.9&-
15.5$\pm$0.7\\
ref-3&0.0000$\pm$0.0007&0.0000$\pm$0.0008&2.3$\pm$0.2&13.8$\pm$1.2&3.4$\pm$1.1 \\
ref-4&-11.0933$\pm$0.0008&-139.6018$\pm$0.0005&3.7$\pm$0.2&-42.3$\pm$0.5&-
43.2$\pm$0.4 \\
ref-5&663.3341$\pm$0.0008&184.9227$\pm$0.0004&4.2$\pm$0.2&6.3$\pm$1.4&-1.6$\pm$1.3 
\\
ref-6&592.0859$\pm$0.0009&104.4975$\pm$0.0006&1.8$\pm$0.1& -13.9$\pm$1.3&-
3.9$\pm$1.2\\
\enddata
\tablenotetext{a}{$RA$ and $DEC$ are positions relative to ref-3 (RA = 
22$^h$52$^m$58\fs56, Dec = -14\arcdeg17\arcmin24\farcs6, J2000.0), epoch = 
1999.592}
\tablenotetext{b}{After adjustment by the model (equations 1 and 2) }
\end{deluxetable}

\begin{center}
\begin{deluxetable}{ll}
\tablecaption{Astrometry of Gl 876 \label{tbl-AST}}
\tablewidth{0pt}
\tablehead{\colhead{Parameter} &  \colhead{Value}}
\startdata
{\it HST} study duration  &1.8 y\\
number of observation sets    &   27 \\
ref. stars $ <V> $ &  $13.6$  \\
\vspace{4pt}
{\it HST} Absolute Parallax\tablenotemark{a}   & 214.6$\pm$0.2   mas\\
{\it HIPPARCOS} Absolute Parallax &212.7 $\pm$2.1 mas\\
YPC95 Absolute Parallax & 211.9 $\pm$ 5.4  mas\\
\vspace{4pt}
{\it HST} Relative Proper Motion\tablenotemark{a}  &1168.3 $\pm$ 1.2 mas y$^{-1}$ 
\\
 \indent in pos. angle & 125\fdg3 $\pm$0\fdg 1 \\
HIPPARCOS Proper Motion  &1174.2 $\pm$ 5.4 mas y$^{-1}$ \\
 \indent in pos angle & 125\fdg1 $\pm$0\fdg6 \\
YPC95 Proper Motion  &1143  mas y$^{-1}$ \\
 \indent in pos. angle & 123\fdg5  \\
\enddata
\tablenotetext{a}{Values come from modeling RV and astrometry simultaneously
(Section~\ref{RVAST})}
\end{deluxetable}
\end{center}

\begin{center}
\begin{deluxetable}{ll}
\tablecaption{Orbital Elements of Perturbation Due to Gl 876b \label{tbl-ORB}}
\tablewidth{0in}
\tablehead{\colhead{Parameter} &  \colhead{Value}}
\startdata
$\alpha$(mas)&0.25 $\pm$ 0.06\\
$\alpha$ (AU) & 0.0012 $\pm$ 0.0003  \\
$i$& 84\arcdeg $\pm$ 6\arcdeg \\
P(days)\tablenotemark{a}& 61.02 $\pm$ 0.03\\
T$_0$ (JD)\tablenotemark{a} & 2450107.87 $\pm$ 1.9\\
e\tablenotemark{a}& 0.10 $\pm$ 0.02\\
$\Omega$ &25\arcdeg $\pm$ 4\arcdeg \\
$\omega$\tablenotemark{a}&338\fdg96 $\pm$ 0\fdg36 \\
$K_1$ (\kms)\tablenotemark{a}&0.210 $\pm$ 0.005 \\
\vspace{4pt}
$M_*$ ($M_{\sun}$)& 0.32$\pm$0.05\\
$M_b$ ($M_{Jup}$)& 1.89$\pm$0.34\\
$M_b$ ($M_{Jup}$)& 1.9$\pm$0.5\tablenotemark{b}\\

\enddata
\tablenotetext{a}{Constrained to values determined from RV measurements.}
\tablenotetext{b}{Error includes $M_*$ uncertainty.}
\end{deluxetable}
\end{center}

\clearpage

\end{document}